\documentclass[11pt]{article}
\pdfoutput=1
\usepackage{amsfonts,amsmath}
\usepackage{amssymb}
\usepackage{fancyhdr}
\usepackage{slashed}
\usepackage{graphicx}
\usepackage{subfigure}
\usepackage{color}

\usepackage{hyperref}
\usepackage[utf8]{inputenc}
\usepackage[titletoc]{appendix}

\def\hybrid{\topmargin -20pt    \oddsidemargin 0pt
        \headheight 0pt \headsep 0pt
        \textwidth 6.25in       % A4 paper
        \textheight 9.25in       % A4 paper
        \marginparwidth .875in
        \parskip 5pt plus 1pt   \jot = 1.5ex}

\hybrid

\def\baselinestretch{1.2}

                                                                                                                                                                                                                                                                                            \catcode`\@=11

\def\marginnote#1{}
\newcount\hour
\newcount\minute
\newtoks\amorpm
\hour=\time\divide\hour by60
\minute=\time{\multiply\hour by60 \global\advance\minute by-\hour}
\edef\standardtime{{\ifnum\hour<12 \global\amorpm={am}%
        \else\global\amorpm={pm}\advance\hour by-12 \fi
        \ifnum\hour=0 \hour=12 \fi
        \number\hour:\ifnum\minute<10 0\fi\number\minute\the\amorpm}}
\edef\militarytime{\number\hour:\ifnum\minute<10 0\fi\number\minute}
\def\draftlabel#1{{\@bsphack\if@filesw {\let\thepage\relax
   \xdef\@gtempa{\write\@auxout{\string
      \newlabel{#1}{{\@currentlabel}{\thepage}}}}}\@gtempa
   \if@nobreak \ifvmode\nobreak\fi\fi\fi\@esphack}
        \gdef\@eqnlabel{#1}}
\def\@eqnlabel{}
\def\@vacuum{}
\def\draftmarginnote#1{\marginpar{\raggedright\scriptsize\tt#1}}

\def\draft{\oddsidemargin -.5truein
        \def\@oddfoot{\sl preliminary draft \hfil
        \rm\thepage\hfil\sl\today\quad\militarytime}
        \let\@evenfoot\@oddfoot \overfullrule 3pt
        \let\label=\draftlabel
        \let\marginnote=\draftmarginnote
   \def\@eqnnum{(\theequation)\rlap{\kern\marginparsep\tt\@eqnlabel}%
\global\let\@eqnlabel\@vacuum}  }

\def\preprint{\twocolumn\sloppy\flushbottom\parindent 2em
        \leftmargini 2em\leftmarginv .5em\leftmarginvi .5em
        \oddsidemargin -.5in    \evensidemargin -.5in
        \columnsep .4in \footheight 0pt
        \textwidth 10.in        \topmargin  -.4in
        \headheight 12pt \topskip .4in
        \textheight 6.9in \footskip 0pt
        \def\@oddhead{\thepage\hfil\addtocounter{page}{1}\thepage}
        \let\@evenhead\@oddhead \def\@oddfoot{} \def\@evenfoot{} }

\def\numberbysection{\@addtoreset{equation}{section}
        \def\theequation{\thesection.\arabic{equation}}}

\def\underline#1{\relax\ifmmode\@@underline#1\else
        $\@@underline{\hbox{#1}}$\relax\fi}

\def\titlepage{\@restonecolfalse\if@twocolumn\@restonecoltrue\onecolumn
     \else \newpage \fi \thispagestyle{empty}\c@page\z@
        \def\thefootnote{\fnsymbol{footnote}} }

\def\endtitlepage{\if@restonecol\twocolumn \else \newpage \fi
        \def\thefootnote{\arabic{footnote}}
        \setcounter{footnote}{0}}  %\c@footnote\z@ }

\catcode`@=12
\relax

\def\figcap{\section*{Figure Captions\markboth
        {FIGURECAPTIONS}{FIGURECAPTIONS}}\list
        {Figure \arabic{enumi}:\hfill}{\settowidth\labelwidth{Figure
999:}
        \leftmargin\labelwidth
        \advance\leftmargin\labelsep\usecounter{enumi}}}
 \relax
\def\tablecap{\section*{Table Captions\markboth
        {TABLECAPTIONS}{TABLECAPTIONS}}\list
        {Table \arabic{enumi}:\hfill}{\settowidth\labelwidth{Table
999:}
        \leftmargin\labelwidth
        \advance\leftmargin\labelsep\usecounter{enumi}}}
 \relax
\def\reflist{\section*{References\markboth
        {REFLIST}{REFLIST}}\list
        {[\arabic{enumi}]\hfill}{\settowidth\labelwidth{[999]}
        \leftmargin\labelwidth
        \advance\leftmargin\labelsep\usecounter{enumi}}}
 \relax
\makeatletter
\newcounter{pubctr}
\def\publist{\@ifnextchar[{\@publist}{\@@publist}}
\def\@publist[#1]{\list
        {[\arabic{pubctr}]\hfill}{\settowidth\labelwidth{[999]}
        \leftmargin\labelwidth
        \advance\leftmargin\labelsep
        \@nmbrlisttrue\def\@listctr{pubctr}
        \setcounter{pubctr}{#1}\addtocounter{pubctr}{-1}}}
\def\@@publist{\list
        {[\arabic{pubctr}]\hfill}{\settowidth\labelwidth{[999]}
        \leftmargin\labelwidth
        \advance\leftmargin\labelsep
        \@nmbrlisttrue\def\@listctr{pubctr}}}
 \relax
\makeatother
\newskip\humongous \humongous=0pt plus 1000pt minus 1000pt

\newif\ifdtup

\relax

\def\be{\begin{equation}}
\def\ee{\end{equation}}
\def\ba{\begin{eqnarray}}
\def\ea{\end{eqnarray}}

\def\del{\partial}

\def\a{\alpha}

\def\b{\beta}

\def\g{\gamma}
\def\G{\Gamma}

\def\e{\epsilon}

\def\th{\theta}

\def\m{\mu}
\def\n{\nu}

  \def\cF{{\cal F}}

\newcommand{\vev}[1]{{\left< {#1} \right>}}

\newcommand{\prt}[1]{{\left( {#1} \right)}}
\newcommand{\prtt}[1]{{\left[ {#1} \right]}}

\def\no{\noindent}

\def\IR{\relax{\rm I\kern-.18em R}}

\def\pp{\partial}

\def\IR{\relax{\rm I\kern-.18em R}}
\def\IL{\relax{\rm I\kern-.18em L}}

\def\inv{^{\raise.15ex\hbox{${\scriptscriptstyle -}$}\kern-.05em 1}}

\def\bea{\begin{eqnarray}}
\def\eea{\end{eqnarray}}
\newcommand{\eq}[1]{(\ref{#1})}
\def\nn{\nonumber}
\def\del{\partial}
\newcommand{\la}[1]{\label{#1}}

\def\a{\alpha}      
\def\b{\beta}       
\def\g{\gamma}  \def\G{\Gamma}  
    
\def\e{\epsilon}

\def\m{\mu} \def\n{\nu}

\def\th{\theta}

\definecolor{markcolor2}{rgb}{1,0,0}

\definecolor{markcolor3}{rgb}{0,1,0}

\newcommand{\half}{\frac{1}{2}}

\begin{document}

\renewcommand{\theequation}{\thesection.\arabic{equation}}
\csname @addtoreset\endcsname{equation}{section}

\newcommand{\beq}{\begin{equation}}
\newcommand{\eeq}[1]{\label{#1}\end{equation}}
\newcommand{\ber}{\begin{eqnarray}}
\newcommand{\eer}[1]{\label{#1}\end{eqnarray}}
\newcommand{\eqn}[1]{(\ref{#1})}
\begin{titlepage}
\begin{flushright}
\hfill{ NCTS-TH/1904}
\end{flushright}
\hfill

\begin{center}

~
\vskip 0.7 cm

{\Large
  \bf
  $c$-Theorem for Anisotropic RG Flows from\\
  Holographic Entanglement Entropy
}

\vskip 0.4in

 {\bf Chong-Sun Chu${}^1$, Dimitrios Giataganas${}^{2,1}$}
 \vskip 0.1in
 {\em
   ${}^1$  Physics Division, National Center for Theoretical
   Sciences, \\
  National Tsing-Hua University, Hsinchu, 30013, Taiwan  \\
\vspace{0.2cm}${}^2$ Department of Physics, University of Athens,
\\University Campus, Zographou, 157 84, Greece
 \\\vskip .1in
 {\tt cschu@phys.nthu.edu.tw, dimitrios.giataganas@phys.uoa.gr
 }\\
 }

\vskip .2in
\end{center}

\vskip .6in

\centerline{\bf Abstract}

We propose a candidate $c$-function in arbitrary dimensional quantum field theories
with broken Lorentz and rotational symmetry.
For holographic
theories we derive the necessary and sufficient conditions on the
geometric background for these $c$-functions to satisfy the $c$-theorem.
We obtain the
null energy conditions for anisotropic background to show that do not
themselves assure the $c$-theorem. By employing them, we find that is possible to impose conditions
on the UV data that are enough to guarantee
at least one monotonic $c$-function along the RG flow.
These UV conditions can be used as building blocks for the
construction of anisotropic monotonic RG flows. Finally,
we apply our results to several known anisotropic
theories and identify the region in the parameters space of the metric
where the $c$-theorem holds for our proposed $c$-function.

\no
\end{titlepage}
\vfill
\eject

%\end{center}

\noindent

\def\baselinestretch{1.2}
\baselineskip 19 pt
\noindent

%%%%%%%%%%%%%%%

\setcounter{equation}{0}

\section{Introduction }

The renormalization group (RG) is a powerful method for constructing
relations between theories at different length scales. It's existence
is fundamental to the explanation of the universality of critical
phenomena. Exact general results for RG flows are important as they
may provide valuable nonperturbative information of strongly coupled
system. The $c$-theorem of Zamolodchikov \cite{Zamolodchikov:1986gt}
is a remarkable result of this kind. It states, for two dimensional
QFTs, the existence of a positive real function $c$ that decreases
monotonically along the RG flow from the ultraviolet (UV) to the
infrared (IR). The function is stationary at the fixed point of the RG
flow, with value given by the central charge of the conformal field
theory (CFT).  The generalization of the two-dimensional $c$-theorem
to higher dimensions was conjectured by Cardy \cite{Cardy:1988cwa} to
hold for any QFT in even dimensions, with the $c$-function given by
the anomaly coefficient associated with the $A$-type Euler density
anomaly. For Lorentz invariant theories in four dimensions, Cardy's
conjecture has been   proven for four dimensions
\cite{Komargodski:2011vj,Komargodski:2011xv}, although the proof
cannot be easily generalized to higher dimensions. However, the
relativistic $c$-theorems have been reformulated by using the
gauge/gravity correspondence
\cite{Girardello:1998pd,Freedman:1999gp,Myers:2010tj,Myers:2010xs}
where monotonic $c$-functions have been proposed in arbitrary
dimensions.

In the Wilsonian formulation of renormalization, the renormalization
group flow is obtained when high energy degrees of freedom are
integrated out and removed from the description. As the quantum
entanglement provides a useful measure of the quantum information
aspects of the theory, quantities related to the entanglement entropy
appear to be natural candidates for the $c$-function. Indeed, for two
dimensional QFT, Casini and Huerta \cite{Casini:2004bw,Casini:2006es}
were able to prove the $c$-theorem by employing the $c$-function
\be
\label{c-2d} c :=3 l \frac{\del S}{\del l}~,
\ee
where $S$ is the
entanglement entropy of a stripe
of length $l$. Properties such as
the subadditivity of the entanglement entropy, the Lorentz symmetry
and unitarity of the QFT were enough to provide the proof of the
$c$-theorem in this framework. The entanglement entropic construction
of the $c$-function immediately suggests a straightforward generalization to
higher dimensions that would provide an intuitive understanding of the
$c$-theorem in terms of the renormalization property of the
entanglement entropy. While a direct study of the renormalization
group property of entanglement entropy is an involved problem in QFT,
the use of AdS/CFT duality transforms it to a much more
tractable one due
to the availability of the Ryu-Takayanagi formula for the entanglement
entropy \cite{Ryu:2006bv}. Indeed the holographic
entropic $c$-function has been shown to obey the $c$-theorem for
Lorentz invariant QFT
\cite{Ryu:2006ef,Myers:2012ed}. Interesting studies along this direction on
RG flows include also
\cite{Albash:2011nq,Liu:2012eea,Park:2018ebm,Kolekar:2018chf}.

Interestingly enough the validity for an entropic $c$--function in theories that exhibit
Lorentz violation has been questioned. There is evidence from the weak coupling analysis
that the entanglement entropy does not decrease monotonically under RG flow
\cite{Swingle:2013zla}. The breakdown of the candidate $c$-theorem
has also been
revealed for holographic Lorentz-violating QFTs
\cite{Cremonini:2013ipa}.
Since broken Lorentz symmetry and broken rotational symmetry are not
uncommon in many condensed matter systems, the study of the
entanglement RG flow under breaking of Lorentz or/and rotational
symmetry is interesting and important.
One of the main goals
of this work is to examine the status of the $c$-theorem in anisotropic QFTs.

In this paper, we will adopt a definition of the $c$-function similar
to the
one in \eq{c-2d}. However it is important to note that in anisotropic theories,  one can in principle have
many different $c$-functions since one can place the stripe in the different
anisotropic directions where the behaviour of the entanglement entropy could be different for each one of them. We remark that in relativistic
holographic CFTs, the monotonicity of the $c$-function is assured as
long as the null energy conditions are satisfied for the bulk gravity
theory \cite{Myers:2012ed}. This is not necessarily the case for anisotropic or Lorentz violating holographic QFT. We show that for anisotropic theories, the
monotonicity of the $c$-function depends also on the satisfaction of
certain geometric background conditions which are complementary to the
null energy conditions. Interestingly there exist UV and/or IR boundary
conditions for generic anisotropic RG flows which when imposed are
enough to guarantee
the existence of at least one monotonically decreasing
$c$-function
along the entire anisotropic flow. Our
result offers new perspectives on
the extension of the application of the $c$-function from isotropic
to anisotropic theories, with potential application to
systems admitting such broken symmetry.

The paper is organized as follows: In section \ref{sec:c}, we propose
a definition of the $c$-function for anisotropic theories.
Then for a generic RG flow, we evaluate the derivative of the proposed $c$-function in terms of the geometric quantities of the background and show that the validity of the $c$-theorem implies an integrated condition on the metric fields.  In section \ref{sec:null}
we derive the null energy conditions and we discuss how they depend on
the way that the rotational symmetry is broken. In section
\ref{sec:suf} we show that the null energy conditions are not enough
to guarantee the monotonicity of the proposed $c$-function under RG
flow. We also derive a set of sufficient conditions, expressed in terms of local properties of the metric, that when satisfied would
assure a monotonically decreased $c$-function. In section \ref{sec:uv}
we show the existence of the UV/IR conditions on the boundary, which when
satisfied
would guarantee the monotonicity
of the $c$-function along the
entire RG flow.  In section \ref{sec:appl}, we apply our analysis to theories with Lifshitz-like anisotropy and to theories with anisotropic hyperscaling violation, and identify regime of parameters where the $c$-theorem holds. In section
\ref{sec:exact}, we discuss the necessary conditions for well behaved
RG flows,  i.e. flows that satisfy the $c$-theorem.
This analysis provides
further evidence on the
definition of our anisotropic $c$-function.  We conclude with a
discussion of
our
results in section \ref{sec:disc}.

\section{A $c$-Function for Theories with Broken Spacetime Symmetry} \la{sec:c}

To motivate the definition of a candidate $c$-function for anisotropic
theory, let us first recall the situation in the standard QFT with Lorentz symmetry.
In two dimensional Lorentzian QFT, the $c$-theorem has been
proven for the $c$-function \eq{c-2d}
where $S$ is the entanglement entropy of an interval of length $l$
\cite{Casini:2004bw,Casini:2006es}.
For a Lorentzian quantum field theory in higher $d$-dimensional space,
\cite{Myers:2012ed} proposed to  consider a 'slab' geometry whose
entangling surface consists of two parallel flat $(d-1)$-dimensional planes
separated by a distance $l$ in flat spacetime metric. For CFT, it is known that
the entanglement entropy for the slab takes
the simple form \cite{Ryu:2006bv,Ryu:2006ef}
\be S_{\rm CFT}=\a
 \frac{H^{d-1}}{\e^{d-1}}-\frac{1}{\b
   \prt{d-1}}C\frac{H^{d-1}}{l^{d-1}}~,
 \ee
 where $\a$ and $\b$ are dimensionless quantities that depend on the spacetime dimension, $\e$ is a UV cutoff and  $H \gg l$
is an infrared regulator for the large distance along the entangling
surface. The second term is proportional to
the central charge $C$ and, by natural induction of \eq{c-2d}, it has been proposed
as a candidate for a $c$-function along the RG flow the function
\cite{Ryu:2006ef,Myers:2012ed}
\be\la{cfunction0}
 c:=\b\frac{l^{d}}{H^{d-1}}\frac{\partial S}{\partial l}~.
 \ee
The $c$-function \eq{cfunction0}
has been shown to satisfy the $c$-theorem for RG flows of
Lorentz invariant holographic
 theories \cite{Myers:2012ed}.

 In this paper, we are interested in theories that admit Lorentz violation
 and anisotropy.
Without loss of generality, let us consider
$d$-dimensional anisotropic space where the rotational symmetry group is
broken down to $SO(d_1) \times SO(d_2)$, with $d_1+ d_2 = d$. Let us
denote the two sub-factors of isotropic space  by $\vec{x}$ and $\vec{y}$.
 The presence of more than two isotropic factors can be treated
 similarly.
 By definition, the metric of a space is of length dimension two and
 in general,
 time coordinate $t$ and spatial coordinates $\prt{\vec{x}, \vec{y}}$
 can be of different length dimensions and they can be
 parametrized as
 \be
[t] = L^{n_t}, \quad [x_i] = L^{n_1}, \quad [y_j] = L^{n_2}~.
 \ee
 For isotropic space, it is $n_1 = n_2$ and for Lorentz invariant space, we have additionally $n_t=n_1$.
Let us now  consider a slab geometry with its entangling surfaces
separated by distance $l_x$ in the $x$-direction and let
$H_x, H_y\gg l_x,l_y$
be the large distances regulating the
infinity along the entangling surface
in the $\vec{x}, \vec{y}$ directions.
We propose the following natural generalization for
the $c$-function in anisotropic theories,
\be
\la{cfunction-x}
c_x := \b_x \frac{l_x^{d_x-1}} {H_x^{d_1-1} H_y^{d_2}}
\frac{\partial S_x}{\partial \ln l_x}~,
\ee
where $\b_x$ is a constant of normalization
that is not important for the current study, $S_x$ is the entanglement
entropy of the stripe with it's complement,
and
\be
d_x := d_1 + d_2 \frac{n_2}{n_1}
\ee
is the effective number of the spatial dimension measured with respect to
the spatial coordinate $\vec{x}$. The appearance of $d_x$ in \eq{cfunction-x}
can be understood since the following dimensionless
combination
$\big(\frac{l_x}{H_x}\big)^{d_1-1} \big(\frac{l_x^{n_2/n_1}}{
  H_y}\big)^{d_2}$
is needed on the right hand side of the expression in order for the
definition of
$c_x$ to be a dimensionless quantity.
In a similar way, an independent $c$-function can be defined for a slab geometry
placed in the $y$-direction
\be
\la{cfunction-y}
c_y := \b_y \frac{l_y^{d_y-1}} {H_x^{d_1} H_y^{d_2 -1}}
\frac{\partial S_y}{\partial \ln l_y},
\ee
where
\be
d_y:= d_1 \frac{n_1}{n_2} +d_2
\ee
is the effective number of spatial dimension measured with respect to
the spatial coordinate $\vec{y}$. The existence of independent $c$-functions
is natural
in anisotropic quantum theory.
In fact, in general there are more
curvature invariants that are consistent with the symmetries of the anisotropic
theory
and hence can appear in the Weyl anomaly. Each of such invariants
is accompanied by a central charge and in principle corresponds to a $c$-function.
We note that for isotropic theories $n_1=n_2$ so $c_x=c_y$ and we recover the single $c$-function \eq{cfunction0}.

Our definition
of the $c$-function
is also motivated by the holographic framework,
where the most generic dual space-time metric for such an anisotropic theory is
\be\la{metric1}
ds_{d+2}^2=-e^{2 B(r)}dt^2+dr^2+e^{2 A_1(r)}dx_i^2+e^{2 A_2(r)}dy_i^2~.
\ee
The space is $d$-dimensional, the $x_i$ coordinates extend along $d_1$
dimensions and the $y_i$ coordinates describe $d_2$ dimensions, such that
$d_1+d_2=d$. The boundary of the space is taken to be at $r\to \infty$.
Then the parameter $d_x$ of the $c$-function definition
\eq{cfunction-x},
is holographically defined at the fixed points. Here we choose to
define
it with respect to the UV fixed point, alternatively it could have
been defined with respect the IR fixed point producing different $d_x$.
The parameter $d_x$ takes the form
\be \la{dx1}
\qquad d_x:= d_1 + d_2 \a~,\qquad \a:= \lim_{r \to \infty} \frac{A_2(r)}{A_1(r)}~,
\ee
which is the total scaling of the boundary spatial system relative to
that of the $x$-direction.

In the following we provide  holographic monotonicity conditions of
the $c$-function  in terms of the bulk metric.
For presentation purpose, we normalize the
entangling action by setting
$4 G_N^{(d+2)}$   as well as the constant $\b_x$ in \eq{cfunction-x}
to unit. We restore the
normalization of the constants
in the final results presented at the end of this
section. The entanglement entropy for the strip with length $l_x$ along the
$x$-direction reads
\be\la{sx}
S_x = H_x^{d_1-1} H_y^{d_2}\int_{r_m}^{r_c}dr
e^{k(r)-A_1(r)}\sqrt{1+e^{2 A_1(r)} x'^2}~,
\quad k(r):=d_1 A_1(r) +d_2 A_2(r)~,
\ee
while for the strip with length $l_y$ along the $y$-direction, it reads
\be
S_y=H_x^{d_1} H_y^{d_2-1}\int_{r_m}^{r_c}dr  e^{k(r)-A_2(r)} \sqrt{1+e^{2 A_2(r)} y'^2}~,
\ee
where $r_m$ is the turning point of the minimal surface and $r_c$
is the boundary cut-off.
In general $S_x \neq S_y$, and converge to each other in the
isotropic limit $A_1(r)=A_2(r)$.
In the following let us demonstrate the methodology with $S_x$ and at
the end of the analysis
we present the results for both $S_x$ and $S_y$.
The equations of motion are written as
\be\la{xp}
x'^2=\frac{e^{2 k_m}}{e^{2 A_1(r)}\prt{e^{2 k(r)} - e^{2 k_m}}}~,
\ee
where $k_m=k(r_m)$ is the constant of motion and $A_{1m}:=A_1(r_m)$, $A_{2m}:=A_2(r_m)$.
The interval distance in terms of the turning point $r_m$ of the surface is
\be\la{length1}
l_x=2 \int_{r_m}^{r_c} dr \frac{e^{k_m-A_1(r)}}{\sqrt{e^{2  k(r) }-e^{2  k_m }}}~.
\ee
To compute the central charge in anisotropic theory, we provide here an
alternative derivation to
\cite{Myers:2012ed}.  Since the equation \eq{length1} is in
principle not analytically integrable
and solvable for $r_m(l)$, we use chain rule to compute the derivative
$\pp S/ \pp l$ by computing
the ratio of $\pp S/\pp r_m$ and $\pp l/\pp r_m$.
By substituting the \eq{xp} in \eq{sx} and differentiating we obtain
\be\la{der1}
\frac{1}{H_x^{d_1-1} H_y^{d_2}} \frac{\pp S_x}{\pp r_m} =
-\frac{e^{2k_m-A_{1m}}}
{\sqrt{e^{2  k(r) }-e^{2  k_m }}}\bigg|_{r\to r_m} +  k_m' e^{2k_m}
\int_{r_m}^{r_c}
\frac{e^{2k(r)-A_1(r)}}{\prt{e^{2  k(r) }-e^{2  k_m }}^{3/2}} dr
\ee
and by acting similarly for  \eq{length1} we get
\be\la{der2}
\frac{\pp l_x}{\pp r_m}=-\frac{2 e^{k_m- A_{1m}}}{\sqrt{e^{2  k(r)
    }-e^{2  k_m }}}
\bigg|_{r\to r_m}+ 2 k_m' e^{ k_m} \int_{r_m}^{r_c}\frac{e^{ 2
    k(r)-A_1(r)}}
      {\prt{e^{2  k(r) }-e^{2  k_m }}^{3/2}} dr~.
\ee
This gives the simple result
\be
\frac{1}{H_x^{d_1-1} H_y^{d_2}} \frac{\pp S_x}{\pp l_x}=\frac{1}{2} e^{k_m}~,
\ee
which when combined with \eq{cfunction-x} relates in a compact way the
central function with
the length of the strip and the turning point of the entangling surface
in the bulk
\be
c_x = \half l_x^{d_{x}} e^{k_m}~.
\ee

To study the monotonicity of the function along the RG flow, we compute
its derivative with respect
to the holographic direction
\be\la{derc}
\frac{\partial c_x}{\pp r_m}= \half e^{k_m} l_x^{d_{x}-1}
\prtt{d_{x}\frac{\pp l_x}{\pp r_m}+l_x k_m'}~.
\ee
Let us take $r_c=\infty$ and rewrite the $l_x$ integrand in the
convenient form to integrate by parts
\be
l_x= 2\int_{r_m}^{\infty} dr F(r)\cdot
\frac{e^{\frac{k(r)}{d_{x}}-A_1(r)}}{k'(r)}~,\qquad
F(r):= \frac{e^{k_m-\frac{k(r) }{d_{x}}} k'(r)}{\sqrt{e^{2k(r)}-e^{ 2 k_m}}}~,
\qquad \cF(r)':=F(r)~,
\ee
where the $F$ function
can be integrated and gives
\be
\cF(r)=- i d_{x} e^{-\frac{k(r)}{d_{x}}}~{}_2
F_1\prtt{\half,-\frac{1}{2  d_{x} },
  1-\frac{ 1}{2 d_{x} }, e^{2 k(r)-2k_m}}~.
\ee
Its derivative with respect to $r_m$ reads
\be
\frac{\pp \cF(r)}{\pp r_m}=- \frac{ k_m' e^{k_m-\frac{k(r)}{d_{x} }}}
     {\sqrt{e^{2k(r)}-e^{ 2 k_m}}}-\frac{k_m'}{d_{x} }\cF(r)~.
\ee
On the other hand, the integration by parts of $l_x$ leads to
\bea\nn
\frac{l_x}{2}=&&\frac{\cF_\infty
  e^{\frac{k_\infty}{d_x}-A_{1\infty}}}{k_\infty'}+
\frac{i \sqrt{\pi} d_{x}  e^{-A_{1m}}}{k_m'}\frac{\G\prtt{1-\frac{1}{2
      d_{x} }}}
     { \G\prtt{\half-\frac{1}{2 d_{x}  }}}
     -\int_{r_m}^{\infty} dr \frac{e^{\frac{k(r) }{d_{x}
         }-A_1(r)}} {k'(r)} E(r)\cF(r)~,
     \ea
     where
$E(r):= \frac{k'(r) }{d_{x} }-A_1'(r)-\frac{k''(r)}{k'(r)}$ and
we have used
$\cF_m=-{i \sqrt{\pi} d_{x}  e^{-\frac{k_m }{d_{x}
  }}\G\prtt{1-\frac{1}{2  d_{x} }}}/
   {\G\prtt{\half-\frac{1}{2 d_{x} }}}~$.
   The derivative of $l_x$ with respect to the turning point
   in the bulk is
   given by
\bea
\frac{1}{2}\frac{\pp l_x}{\del r_m} &=& \frac{\pp_{r_m}\cF_\infty
  e^{\frac{k_\infty}{d_x}-A_{1\infty}}}
     {k_\infty'}
 - i \sqrt{\pi} e^{-A_{1m}} \frac{\G\prtt{1-\frac{1}{2 d_{x} }}}{
   \G\prtt{\half-\frac{1}{2  d_{x} }}}
 \\
 &+& k_m' e^{k_m} \int_{r_m}^{\infty}
 dr\frac{e^{-A_1(r)}}{k'(r)\sqrt{e^{2  k(r) }-e^{2  k_m }}}  E(r)
+ \frac{k_m'}{d_{x} }\int_{r_m}^{\infty} dr \frac{e^{\frac{k(r)
    }{d_{x} }-A_1(r)}}{k'(r)} E(r) \cF(r). \nn
\eea
By substituting the above expressions to the \eq{derc}, we find that
the derivative
of the $c$-function can be rewritten in a fairly compact form as
\be\la{derx}
\frac{4 G_N^{(d+2)}}{\b_x} \frac{\partial c_x}{\pp r_m}=   e^{k_m}
l_x^{d_{x}-1} d_{x}
\prtt{ k_m' \int_0^{l_x} dx \frac{1}{k'(r)}\prt{\frac{k'(r)}{d_{x} }
    -A_1'(r)-\frac{k''(r)}{k'(r)}}}~,
\ee
where the integral has been expressed with respect to $x$ using the
equation \eq{xp} and
we have restored the
$4 G_N^{(d+2)}/\b_x$
units according to the
discussion after the equation \eq{dx1}. Similarly,
the $c$-function obtained from the strip along the $y$-direction   gives
\be\la{dery}
\frac{4 G_N^{(d+2)}}{\b_y} \frac{\partial c_y}{\pp r_m}=  e^{ k_m}
l_y^{d_{y}-1} d_{y}
\prtt{ k_m' \int_0^{l_y} dy \frac{1}{k'(r)}\prt{\frac{k'(r)}{d_{y}}
    -A_2'(r)-\frac{k''(r)}{k'(r)}}}~.
\ee
To treat effectively the boundary terms on the above equations, we
have assumed that close to
the boundary the following conditions hold
\be\la{conif}
\lim_{r\to \infty}k'(r)e^{A_{1}(r)+k(r)}=\infty~,\qquad \lim_{r\to \infty}
k'(r)e^{A_{2}(r)+k(r)} =\infty~.
\ee
We remark that these conditions can usually be easily satisfied
since we have
$\lim_{r\to \infty}
e^{A_{i}(r)}=\infty~$ at the spacetime boundary.

\section{The Null Energy Conditions for Anisotropic Theories  } \la{sec:null}

In an isotropic theory, the employment of the null energy conditions (NEC)
ensure non-repulsive gravity \cite{Wald:106274},
and  avoid
instabilities \cite{Buniy:2006xf} and superluminal modes
\cite{Dubovsky:2005xd,Hoyos:2010at} in the scalar correlators of the
theory. We expect the situation would be similar in the anisotropic case
and so we impose the NEC
for the matter fields that drive the holographic RG
flow:
\be \la{neca}
T_{\m\n} \xi^\m \xi^\n\ge0,\qquad \xi^\m \xi_\m=0~,
\ee
where $\xi$ are the null vectors. Eliminating $\xi_0$, the NEC can be written in the form
\be\la{xtmn}
\xi^i{}^2 g_{ii}\prt{T_i^i-T_0^0}+\xi^j{}^2g_{jj}\prt{T_j^j-T_0^0}+
\xi^r{}^2 g_{rr}\prt{T_r^r-T_0^0} \ge0~.
\ee
Here $i$ and $j$ indices denote the $x$ and $y$ directions
respectively and the repeated indices above are summed only one
time. We have used the fact that diagonal anisotropic metric leads to
diagonal anisotropic tensor.  By contracting the Einstein gravity
equations and allowing sources we can rewrite the above equation in
terms of the Ricci tensor since
\be
R_{\m\n} \xi^\m \xi^\n =T_{\m\n}\xi^\m\xi^\n~.
\ee
Note that $\xi^i,~\xi^j$ and $\xi^r$ are independent in  \eq{xtmn}, so the NEC read
\be\la{null1}
 R_i^i-R_0^0\ge 0~,\qquad R_j^j-R_0^0\ge 0~,\qquad R_r^r-R_0^0\ge 0~,
\ee
where no summation takes place. Notice that the number of conditions
increase with the number of subgroups that the rotational symmetry of
the space is broken to.  In our case, we break the rotational symmetry
to a product of two subgroups described by the directions $x$ and $y$,
encoded in the first two independent conditions of \eq{null1}. When
the isotropy is restored the first two conditions become equivalent.

Applying our generic formulas \eq{null1} for the anisotropic spacetime
\eq{metric1} we have three independent NEC,
\bea\la{nec1}
&&g_1'(r) :=\prt{\prt{B'(r)-A_1'(r)} e^{B(r)+k(r)}}'\ge 0~,\\\la{nec2}
&&g_2'(r) :=\prt{\prt{B'(r)-A_2'(r)} e^{B(r)+k(r)}}'\ge0~,\\\la{nec3}
&&N_3 :=  -d_1 A_1'(r)^2-d_2 A_2'(r)^2+ B'(r) k'(r)-k''(r)\ge 0~.
\eea
where the first two can be written in terms of the monotonically
increasing functions $g_1(r)$ and $g_2(r)$.
The third condition can be written as
\be\la{eqfa}
f'(r)e^{-k(r)/\prt{d_1+d_2}+B(r)}-\frac{d_1 d_2}{d_1+d_2} \prt{A_1'(r)-A_2'(r)}^2 \ge 0~,
\ee
where
\be\la{deffa}
f(r):=-k'(r) e^{k(r)/\prt{d_1+d_2}-B(r) }~,
\ee
is a monotonically increasing function since $f'(r)\ge 0$. The reason
we rewrite the NEC in terms of monotonic functions is that in certain
cases the boundary data will be enough to ensure the
monotonicity of the $c$-function along the whole RG flow. This will
become clearer in the next sections.

\section{Sufficient Conditions of Monotonicity of
  the $c$-function} \la{sec:suf}

It is convenient to use the NEC \eq{nec3} to eliminate the
second derivatives of the equations
\eq{derx} and \eq{dery} and  obtain
\bea\la{set2}
\frac{\partial c_x}{\pp r_m}=  \frac{\b_x e^{k_m} l_x^{d_{x}-1} d_{x} }
     {4 G_N^{(d+2)}}k_m'\int_0^l \frac{dx}{k'(r)^2}\prt{N_3 + d_2 A_2'(r)
       \prt{A_2'(r)-A_1'(r)}+ k'(r)\prt{\frac{k'(r)}{d_{x} }-B'(r)}},~
     \\\la{set22}
     \frac{\partial c_y}{\pp r_m}=  \frac{\b_y e^{k_m} l_y^{d_{y}-1} d_{y}  }
          {4 G_N^{(d+2)}}k_m'\int_0^l \frac{dy}{k'(r)^2} \prt{N_3
            + d_1 A_1'(r)\prt{A_1'(r)-A_2'(r)}+ k'(r)\prt{\frac{k'(r)}{d_{y} }
              -B'(r)}}.~
          \eea

\subsection{Sufficient Condition for Anisotropic Theories}

  The necessary condition for the  monotonicity of the functions $c_x$, $c_y$
along the RG flow
is that the positive expression $N_3$ is large enough to compensate
the contributions of the other terms when integrated over the RG flow.
However, the integration needs the knowledge of the explicit form of
the theory. Instead, a set of sufficient conditions of monotonicity
for general theories can be formulated as local conditions in the form
\be\la{sufcon1}
k'(r)\ge0~,
\ee
and
\bea \la{sufcon2}
d_2 A_2'(r)\prt{A_2'(r)-A_1'(r)}+ k'(r)\prt{\frac{k'(r)}{d_{x} }
  -B'(r)}\ge 0~, \quad \mbox{for the monotonicity of $c_x$},\\
\la{sufcon3}
d_1 A_1'(r)\prt{A_1'(r)-A_2'(r)}+ k'(r)\prt{\frac{k'(r)}{d_{y} }
  -B'(r)}\ge 0~,\quad \mbox{for the monotonicity of $c_y$}.
\eea

Note that the condition \eq{sufcon1} implies that $d_1 A_1'(r)\ge - d_2
A_2'(r)$.
This allows for negative $A_1'(r)$ or $A_2'(r)$
as long as their overall sum above is positive. This is in contrast
already to isotropic (non-)relativistic theories where the single
$A'(r)$ has to be positive. Furthermore, we note that backgrounds with
$B'(r)\le0$ and $A_1'(r) A_2'(r)\le0$ always satisfy the set of
constrains as long as $k'(r)\ge0$.
We also note that the above sufficient conditions can be
expressed in terms of monotonic functions as
\bea\la{sufcon2a0}
&& f(r)\le0~,\\
\la{sufcon2a}
&& f'(r) e^{B(r)-\frac{k(r)}{d}}+\frac{f(r) e^{-\frac{(d+1) k(r)}{d}}}{d_1+d_2}
\prt{\prt{d_1 -d_2} g_1(r) +2 d_2 g_2(r)}+ \frac{k'(r)^2 d_2(1-\a)}
    {\prt{d_1+d_2}\prt{d_1+d_2 \a}}\ge0,\;\;\;\;\;\;~~
    \\
    \la{sufcon2b}
    && f'(r) e^{ B(r)-\frac{k(r)}{d}}+\frac{f(r) e^{-\frac{(d+1) k(r)}{d} }}{d_1+d_2}
    \prt{\prt{d_2 - d_1} g_2(r) +2 d_1 g_1(r)}-
    \frac{k'(r)^2 d_1 (1-\a)}{\prt{d_1+d_2}\prt{d_1+d_2 \a}}\ge0, \;\;\;\;\;\;~~
\eea
where $f(r)$ and $g_{1,2}(r)$ are the monotonically increasing
functions given by \eq{deffa}, \eq{nec1} and \eq{nec2} respectively,
and $\a$ is defined in \eq{dx1}. As we will show in the next section,
the form \eq{sufcon2a0}-\eq{sufcon2b} of inequalities are useful for
imposing the UV and IR criteria on the RG flows in order to guarantee
monotonicity of our proposed $c$-functions.

\subsection{Sufficient Condition for Isotropic UV Fixed Points}

For anisotropic RG flows with
isotropic UV fixed point, the symmetries of the
theory are restored in the UV (i.e. $\a =1$) and the
conditions \eq{sufcon2a0}-\eq{sufcon2b} simplify to,
\footnote
    {For the special case of anisotropic theory with equal
      number of anisotropic dimensions $d_1=d_2$, the conditions are
      further simplified to
$g_i(r)\le0, f(r)\le0$.
      }
\be\la{sufconcases}
f(r)\le0~,\qquad \prt{d_1 -d_2} g_1(r) +2 d_2 g_2(r)\le 0~,
\qquad \prt{d_2 - d_1} g_2(r) +2 d_1 g_1(r)\le0~.
\ee
We note that for an isotropic and
non-relativistic RG flow: $A_1(r)=A_2(r)\neq B(r)$ and
$d_{x}=d_{y}=d$, we recover from \eq{sufcon2a0}-\eq{sufcon2b}
the sufficient conditions of
\cite{Cremonini:2013ipa}:
$f(r)\le 0$ and $g_2(r)\le0$, with
$g_1(r)=g_2(r)$.
If the background is also
conformal $B(r)=A_2(r)= A_1(r)$, then we have $d_{x}=d_{y}=d$ and $k(r)=d
A_2(r)$ and the sufficient conditions of monotonicity
follows directly from the NEC without the need of any other condition
on the metric.
Finally we remark that the sufficient conditions presented here are by no mean
unique, but are the ones with minimal number of constraints imposed in addition to the NEC.
We will present in the appendix \ref{app:a1}, for example,
another set of sufficient conditions on the metric which takes a
simpler but more restrictive form.

\section{Asymptotics and Boundary Criteria of Monotonicity} \la{sec:uv}

The conditions \eq{sufcon1} - \eq{sufcon3}, or equivalently,
\eq{sufcon2a0} - \eq{sufcon2b} are conditions imposed on
the metric for all $r$. This corresponds to conditions in the field
theory that have to be imposed for the whole range of energies.
Physically, it is desirable to have a weaker form of conditions. In
this section, we show that
it is possible to replace these bulk
conditions with conditions on the boundary
data such that at least one of the $c$-functions
is monotonic along the entire RG flow.

\subsection{Boundary Condition on the Geometry}\label{bcg}

Let us start with the condition \eq{sufcon2a0}. Due to the NEC \eq{eqfa},
$f' \geq 0$ and so \eq{sufcon2a0} is  guaranteed if the
following boundary condition
\be\la{fuv}
f{}_{UV}\le 0
\ee
is imposed.
As for the condition \eq{sufcon2a} for $\del c_x/\del r_m \geq 0$,
we find that it is guaranteed if the following condition is satisfied:
\be\la{uvconditionsx}
\a \leq 1 \quad \mbox{and}\quad
\begin{cases}
   d_1>d_2: \quad  g_1{}_{UV}, \; g_2{}_{UV} \le 0~, \\
d_1=d_2: \quad    g_2{}_{UV} \le 0~,\\
d_1<d_2: \quad  g_1{}_{IR}\ge 0,~g_2{}_{UV} \le 0~,\\
\end{cases}
\ee
Similarly, the condition \eq{sufcon2b} for $\del c_y/\del r_m \geq 0$
is guaranteed if the following condition is satisfied:
\be\la{uvconditionsy}
\a\geq 1 \quad \mbox{and}\quad
\begin{cases}
   d_1>d_2: \quad g_1{}_{UV}\le 0,~g_2{}_{IR} \ge 0~,\\
  d_1=d_2: \quad   g_1{}_{UV} \le 0~,\\
  d_1<d_2: \quad  g_1{}_{UV}, \; g_2{}_{UV} \le 0~.\\
  \end{cases}
\ee
It is interesting to note that, except for the case of $\a=1$ and $d_1=d_2$, it is
generally impossible to impose the boundary data to satisfy both
\eq{uvconditionsx} and \eq{uvconditionsy} simultaneously. Therefore, what
we have shown is that
it is possible
to impose the conditions \eq{uvconditionsx} or \eq{uvconditionsy}
at the UV or IR such that
the $c$-theorem holds for at least one of our proposed $c$-functions.
This does not preclude the other candidate $c$-function to satisfy
the $c$-theorem. In fact we will show in section \ref{sec:exact} that our proposed
$c$-functions indeed satisfy the $c$-theorem for several known anisotropic backgrounds.

\subsection{UV Criteria on Fefferman-Graham Expansion}

The above boundary conditions \eq{fuv}, \eq{uvconditionsx}, \eq{uvconditionsy}
are in terms of the geometry. We will show next that for asymptotically AdS geometry,
it is possible to formulate the sufficient conditions in terms of boundary
data of the field theory.

Let us implement a UV analysis with an asymptotically AdS geometry.
We employ the standard
Fefferman-Graham (FG) expansion to write the subleading corrections at
the UV
($r \to \infty$) as
\be \label{AAB}
A_i(r)=r+\half \g_{a_i} e^{-b  r}+ \cdots~,\qquad
B(r)=r+\half \g_{b} e^{-b r}+\cdots~, \quad b>0.
\ee
This just means  that $\a =1$ and we are looking at the
conditions \eq{sufconcases} for asymptotically isotropic space.
For a flat boundary in the absence of sources,
we have $b=d$ and the coefficients $\g_{a_i}, \g_{b}$ are the VEV of the
dual stress tensor: $\g_{a_i}\sim \vev{T_{ii}}$,
$\g_{b}\sim \vev{T_{00}}$ and they encode the  breaking of the Lorentz
symmetry and isotropy. In order to acquire the
sufficient conditions for these parameters that ensure the right
monotonicity of the RG flow, we allow in principle additional sources
in the theory. Plugging \eq{AAB} into the definitions of $f$ and
$g_i$, we obtain
\bea
&&f(r)=-\prt{d_1+d_2} +\half e^{-b r}\big(\prt{b-1}
\prt{d_1 \g_{a_1}+d_2 \g_{a_2}}+\prt{d_1+d_2} \gamma_b\big)+\cdots~,
\label{fr}\\
&&g_i(r)= \half b e^{(d- b)r }\prt{\g_{a_i}-\g_b}+\cdots \label{gir}
\eea
near the boundary $r \to \infty$.
Without loss of generality, let us assume that $d_1 \geq d_2$.
Let us first examine the NEC and spells out the constraints on the
background.
The NEC \eq{eqfa} implies the condition
\be \label{nec-bdy}
  \prt{1-b}\prt{d_1 \g_{a_1}+d_2
    \g_{a_2}}-\prt{d_1+d_2} \g_b \ge0~,
  \ee
while the NEC \eq{nec1}, \eq{nec2}  reads
\be \label{ccc}
(d-b) \prt{\g_{a_i}-\g_b} e^{(d-b) r}\ge 0.
\ee
If $b\ge d$,
then it follows from \eq{ccc} that \footnote{
  The case $\g_{a_i} = \g_b$ is more subtle
  as higher order subleading corrections need to be included in the analysis
  and will not be considered here.
}
\be \label{gab}
\g_{a_i} < \g_b~.
\ee
This implies immediately that  $g_{i UV} < 0$
and the  monotonicity condition
\eq{uvconditionsx}   is satisfied for the function $c_x$.
On the other hand, if $b < d$, then \eq{ccc} implies that $\g_{a_i} > \g_b$.
This means $g_{i UV} > 0$ and the monotonicity conditions
\eq{uvconditionsx} cannot be satisfied. Working similarly for $d_1 \leq d_2$,
we  reach  similar conclusions for the $c_y$ function.

To summarise this subsection, we find the conditions for the boundary data
of the field theory
that when satisfied would guarantee a monotonic $c_x$-function or
$c_y$-function. Generic backgrounds which satisfy at the boundary one
of the inequalities \eq{uvconditionsx}, \eq{uvconditionsy}, or in
terms of the Fefferman-Graham expansion, the NEC
\eq{nec-bdy}-\eq{gab}, are guaranteed to have the desirable behavior along
the RG flow for the function $c_x$ or $c_y$. For backgrounds with isotropic UV fixed points and equal number of anisotropic dimensions, appropriate UV conditions can guarantee the  monotonicity for both $c_x$ and $c_y$.

\section{Sufficient Conditions Applied on Certain Anisotropic Theories}  \la{sec:appl}
Let us briefly demonstrate and discuss the sufficient conditions on some
interesting theories.

\subsection{Lifshitz-like Anisotropic Symmetry}

We first demonstrate our methods on the simpler geometries that exhibit
Lifshitz scaling symmetry:
\be\la{lif}
ds^2=-e^{2 z r}dt^2+ e^{2  r}\prt{dx^2+dy^2}+dr^2~,
\ee
where $z$ is the Lifshitz scaling exponent measuring the degree of
Lorentz symmetry violation.
The NEC \eq{nec1}-\eq{nec3} require $z\ge1$, while the monotonicity
conditions \eq{sufcon1}-\eq{sufcon3} give $z\le 1$. So the sufficient
condition is guaranteed only for $z=1$, the AdS space, as also noted in
\cite{Cremonini:2013ipa}.
We note that this is only a sufficient condition.
In fact due to the fact since the exponent functions in the
metric are all linear in $r$, the integrand \eq{derx}  is zero for
all the values for $z\ge 1$ and hence the $c$ function theorem is satisfied.
This is expected from the scale invariance
of the theory and we will elaborate further on this point in the next section
\ref{sec:exact}.

The analysis becomes more interesting when the Lifshitz-like
anisotropic symmetry is present, which is realized by the following metric
\be\la{lifaniso}
ds^2=e^{2 z r}\prt{-dt^2+dx^2}+ e^{2  r} dy^2 +dr^2~,
\ee
where $z$ measures the degree of Lorentz symmetry violation and anisotropy. We  compute the entanglement entropy using the equations \eq{sx} and
\eq{length1} and get
\bea \la{entz}
&&S_{x}= N^2 H_x^{d_1-1} H_y^{d_2}\prtt{\frac{\b_1}
  {\e^{d_1-1+\frac{d_2}{z}}}-\frac{\b_2}{l_x^{d_1-1+\frac{d_2}{z}}}}~,\\
&& S_{y}= N^2 H_x^{d_1} H_y^{d_2-1}\prtt{\frac{\tilde{\b}_1}
  {\e^{d_1 +\frac{d_2-1}{z}}}-\frac{\tilde{\b}_2}{l_y^{d_1 z+(d_2-1)}}}~.
\eea
The constants $\b_i,~\tilde{\b}_i$ depend on  $d_x$ and $d_y$ and are
not important for our discussion.  Notice that the scaling dimensions  $d_{x}=d_1+d_2/z$ and
$d_{y}=d_1 z+ d_2 $  appear in the entanglement entropy formula above.
This demonstrates explicitly
that  the definition \eq{dx1} for $d_x$ and the alike for $d_y$ are appropriate
in our proposed $c$-functions
\eq{cfunction-x}, \eq{cfunction-y}.

The NEC \eq{nec1}-\eq{nec3} give $z\ge1$.  The treatment of the
boundary terms  \eq{conif} give the following conditions when the NEC are applied:
$d_2+z(1+d_1)\ge0$ and $1+d_2+d_1 z\ge0$ which are satisfied
trivially.  Then monotonicity conditions give
\eq{sufcon1}-\eq{sufcon3}
\bea
d_2\prt{1-z}\ge0~,\qquad && d_1   +\frac{d_2}{z}\ge 0~\quad
\eea
Therefore again the sufficient conditions and the NEC together can be satisfied only for $z=1$, i.e. AdS
space-time.

So far our metrics have been restricted to linear exponents, below we
present more general backgrounds.

\subsection{Anisotropic Hyperscaling Violation Symmetry}

Next let us consider isotropic theories that exhibit Lifshitz scaling
and hyperscaling violation as described by the following metric
\be\la{hyp}
ds^2= -  \prt{\frac{\th r}{d}}^{2\prt{1-\frac{z d}{\th}}}  dt^2+
\prt{\frac{\th r}{d}}^{2\prt{1-\frac{d}{\th}}} \prt{dx^2+ dy^2 }+dr^2~,
\ee
where $\theta$ is the hyperscaling violation exponent.  The NEC
\eq{nec1}-\eq{nec3} give
\be \la{lif1}
\prt{z-1 }\prt{d+z-\th}\ge 0~,\qquad
\prt{d-\th}\prt{d (z-1)-\th}\ge 0~,
\ee
while the monotonicity conditions \eq{sufcon1}
and \eq{sufcon2} read:
\be \la{lif2}
1- \frac{d}{\th}\ge 0~, \qquad
\frac{z-1}{\th} \ge 0~.
\ee
As for the condition \eq{conif}, it reads
\be\la{thregion}
\frac{d+1}{\th}\le 1~.
\ee

\begin{figure}[t!]
\begin{minipage}[ht]{0.5\textwidth}
  \begin{flushleft}
\centerline{\includegraphics[width=73mm ]{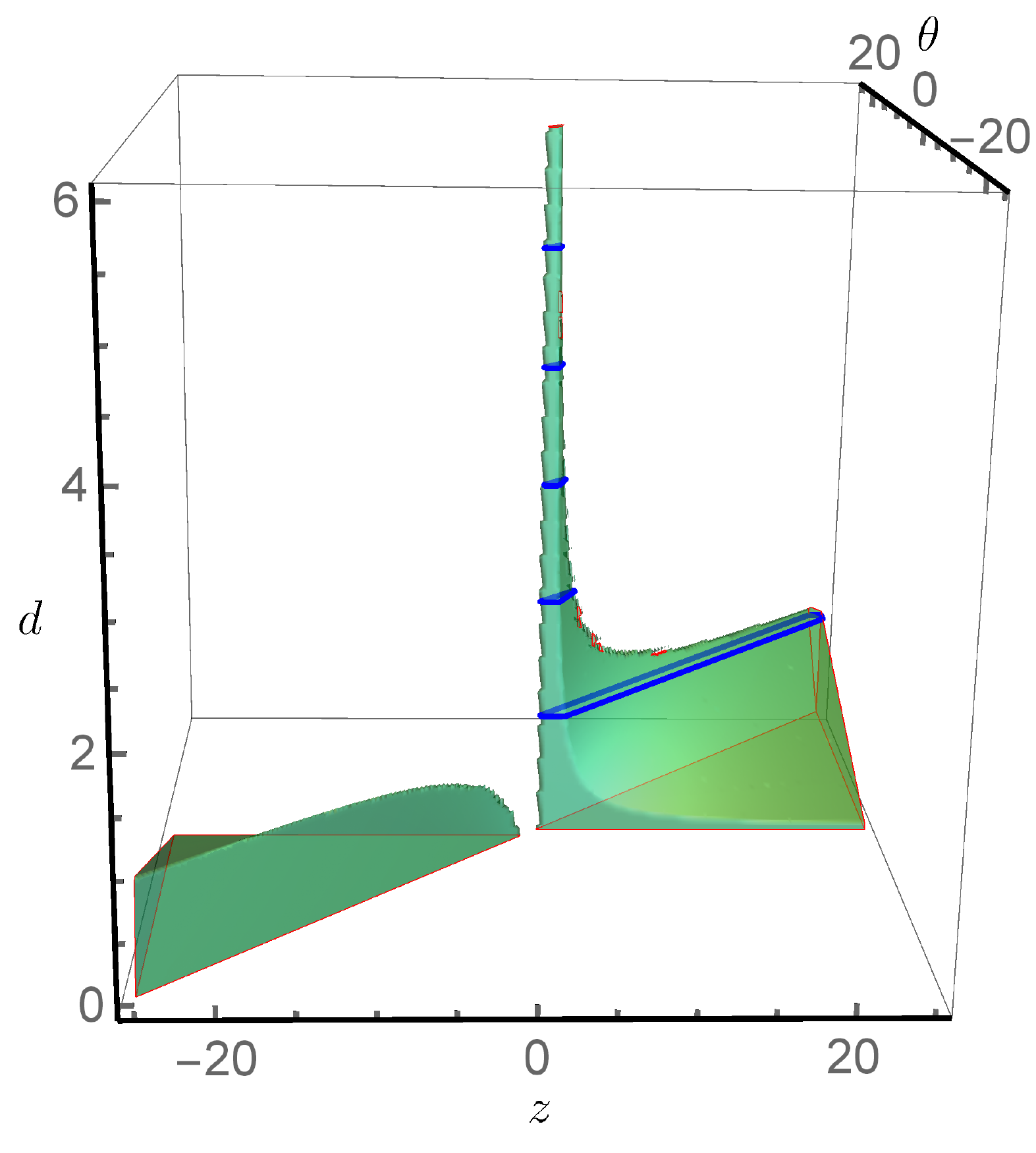}}
\caption{\small{The parametric volume $(\th,~z,~d)$ that is sufficient
    to guarantee RG flows with the right $c$-function
    monotonicity. The meshed lines on the surface represent the values
    of the integer spatial dimensions $d$. Notice how drastically the
    increase of the dimension $d$ shrinks the
    allowed region in the parametric space
    that gives a well behaved RG flow. The necessary
    conditions of section \ref{sec:exact} are more tolerant.}}
\label{fig:hysca1}\vspace{+.0cm}
\end{flushleft}
\end{minipage}
\vspace{-.0cm}
\hspace{0.3cm}
\begin{minipage}[ht]{0.5\textwidth}
\begin{flushleft}
\centerline{\includegraphics[width=77mm]{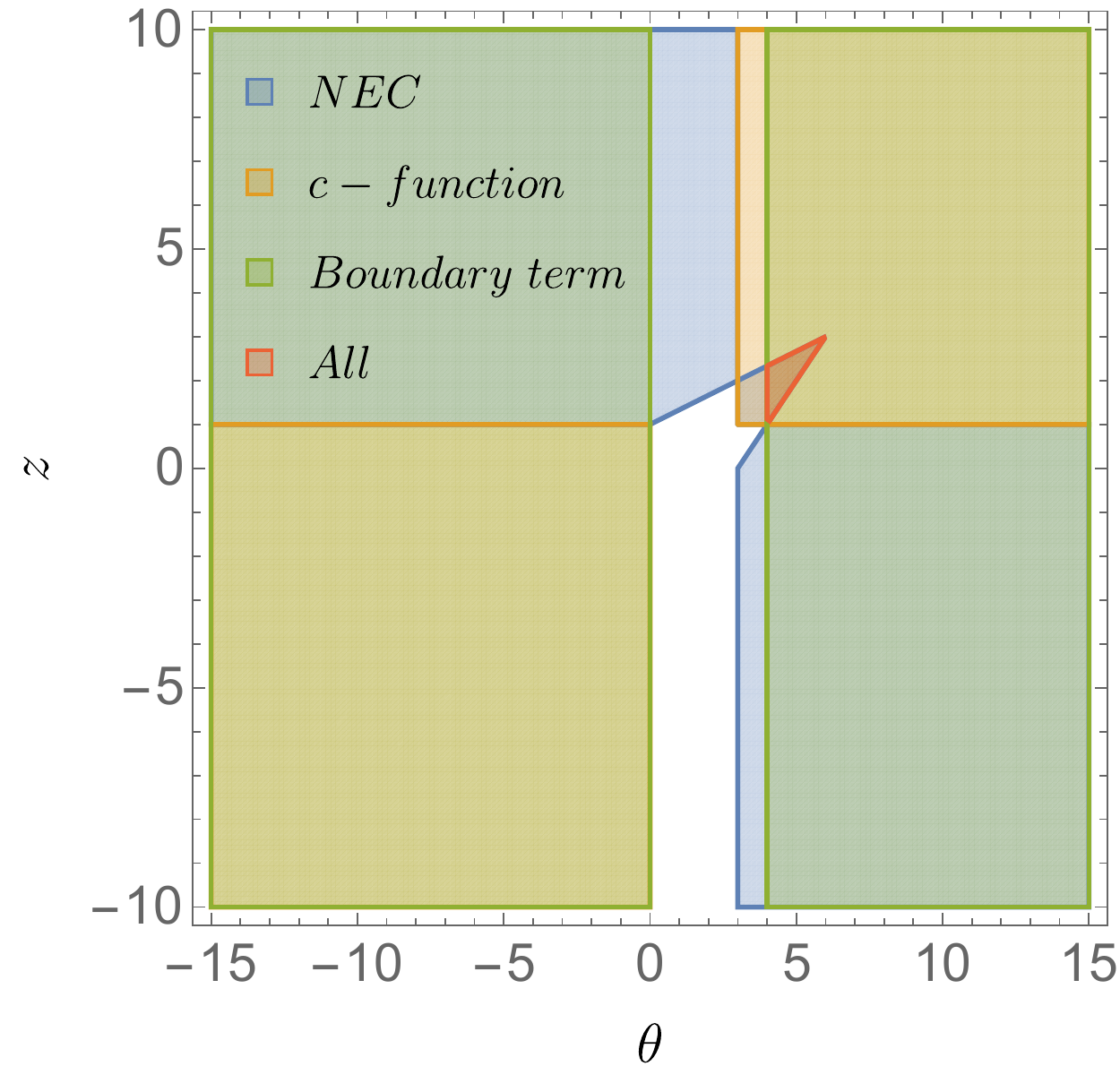}}
\caption{The window of parameters that lead always to well behaved RG
  flows for $d=3$ space-time dimensions. The different coloring
  represents where the different conditions are satisfied and the
  triangle is the common region in the parameter space that is
  sufficient to guarantee well behaved RG flow.}
\label{fig:hysca2}\vspace{+.0cm}
\end{flushleft}
\end{minipage}
\end{figure}

We point out that the allowed region in the parametric space is quite small
and we depict this region in the figures \ref{fig:hysca1} and
\ref{fig:hysca2}. We note that as we increase the number of dimensions,
the sufficient condition becomes more difficult to satisfy.

The sufficient $c$-function conditions for theories that exhibit
anisotropic Lifshitz scaling and hyperscaling violation are more
involved.
These symmetries are described by the following metric
\be\la{hypaniso}
ds^2= -  \prt{\frac{\th r}{d}}^{2\prt{1-\frac{z d}{\th}}}
\prt{-dt^2+dx^2}+ \prt{\frac{ \th r}{d}}^{2\prt{1-\frac{d}{\th}}} dy^2 +dr^2~.
\ee
The NEC \eq{nec1} give
\be \la{cc1}
\prt{z-1}\prt{d_2-\th+z\prt{1+d_1}}\ge 0~,\qquad
 \th\prt{\th-z d}+d_2 d \prt{z-1}\ge 0~,
\ee
  while the monotonicity condition \eq{sufcon1} gives
  \be\la{cc2a}
  1-\frac{d_2 + d_1 z}{\th} \ge 0,
  \ee
  and the monotonicity conditions \eq{sufcon2} and \eq{sufcon3} give the
  same condition
  \be \la{cc2b}
 (\frac{z-1}{\th}) (1- \frac{d}{\th}) \ge 0.
\ee
The elimination of the boundary terms \eq{conif} provide extra conditions
\be\la{cc3}
 1-\frac{d_2+z+d_1 z}{\th} \ge0~,\qquad   1-\frac{1+d_2+d_1 z}{\th} \ge0~.
\ee
We have confirmed that the
conditions \eq{cc1}, \eq{cc2a},
\eq{cc2b} and \eq{cc3} are compatible with each other and
they can be satisfied simultaneously, for a small region in
the parametric space, as shown in figures \ref{fig:hysca1} and \ref{fig:hysca2}.

As seen from the form \eq{hypaniso} of the metric,
the isotropic limit is obtained by setting
$d_1= 0$ and $d_2=d$. Indeed by doing so and  ignoring all the
conditions derived corresponding to the $x_i$ dimensions,
the conditions \eq{cc1}, (\ref{cc2a}, \ref{cc2b})
and the second condition in \eq{cc3} become
\eq{lif1}, \eq{lif2}, \eq{thregion} respectively for the Lifshitz metric.

We note that our analysis
can be applied to the vacuum anisotropic hyperscaling violation theory
which has the same background form \eq{hyp}, derived
by a generalized Einstein-Axion-Dilaton action
\cite{Giataganas:2017koz} with a potential proportional to the
exponential of the dilaton \cite{Giataganas:2018ekxv}.
It can be checked easily that
this background satisfies our stated sufficient conditions and hence
its RG flows satisfy the $c$-theorem.

\section{Necessary Conditions on Anisotropic RG flows } \la{sec:exact}

So far we have considered the sufficient conditions for monotonic RG flows.
In this section by performing explicitly the integrations of
\eq{derx} and \eq{dery}, we confirm that the necessary conditions are indeed much
more relaxed. The $c$-function integral can be written as
\be\la{derxexact}
\frac{4 G_N^{(d+2)}}{\b_x} \frac{\partial c_x}{\pp r_m}=   e^{k_m} l_x^{d_{x}-1} d_{x}
\prtt{ k_m' \int_{r_m}^{\infty} dr \frac{e^{k_m-A_1(r)}}{k'(r)^2\sqrt{e^{2k(r)}-e^{2km}}}
  \prt{\frac{k'(r)^2}{d_{x} }-A_1'(r)k'(r)-k''(r)}}
\ee
and a similar expression exists for the $c_y$ according to \eq{dery}.

In the previous section, we have checked that for the Lifshitz space \eq{lif}
to give rise to a monotonic RG-flow, the sufficient condition allows only the
case of $z=1$. However, since the Lifshitz space has a scaling symmetry,
one can expect the dual field theory to have a conformal symmetry and so the
$c$-function should not run at all. Indeed we have checked analytically that the
integrand of \eq{derxexact} vanishes identically for any value of $z$. Therefore
as long as the background is allowed by the NEC, i.e. $z\geq 1$, the
dual theory
has constant $c$-functions and satisfy the $c$-theorem.

For the same reason one can expect that the same holds for the
Lifshitz-like anisotropic spaces \eq{lifaniso}. We have checked this analytically
and again the integrand cancel out completely for any value of $z$. It is
satisfying to see that this cancellation
takes place only because of our choice of $d_x$ and $d_y$ in the definition of
the $c$-function.
As an example, the field theories with space-dependent $\theta$-term
coupling, which have a Lifshitz-like anisotropic symmetry with $z=3/2$
\cite{Azeyanagi:2009pr}
will satisfy the $c$-theorem for our choice of
the
$c$-functions.

For general background with more complicated metric, the analytic method cannot
be applied and one has to resort to numerical method integrating the equation \eq{derxexact}. Generically the analysis is complicated and the sufficient condition obtained from the study conducted in this work
could provide valuable insights on
what region in the parameters space one should start focusing on.

\section{Discussion} \la{sec:disc}

Motivated by the relevant
construction in two-dimensional quantum field theories
\cite{Casini:2004bw,Casini:2006es}, we have constructed an extension of the $c$-function \eq{cfunction-x}
for higher dimensional anisotropic theories. Our proposal suggests
the presence of as many independent $c$-functions as the number of independent
isotropic factors within the anisotropic geometry; and that
they would become at the IR fixed point, the central charges of the
underlying (an)isotropic theories. Our proposed $c$-function  relies on
the knowledge of the entanglement entropy of a strip-shaped region and
the relative scaling between the spatial directions at the fixed
points. It has no UV divergences, although the entanglement entropy is
divergent
itself. When the full rotational symmetry is restored, our
$c$-functions converge and reduce to the original proposal
\cite{Ryu:2006ef,Myers:2012ed} of the $c$-function for the
isotropic case.

With the use of the null energy conditions, the sufficient conditions
ensuring a decreasing $c$-function towards the IR take the form
\eq{sufcon1}-\eq{sufcon3}.
These conditions
can be expressed in terms of monotonic
functions as \eq{sufcon2a0}-\eq{sufcon2b}. In the case the anisotropic flow
admit isotropic UV fixed points, these conditions
reduce to \eq{sufconcases}. We point
out that the null energy conditions in general do not guarantee a well behaved
anisotropic RG flow, in contrary to what happens in conformal
theories.

We have also derived the necessary conditions for the right monotonicity of the $c$-function. They are  expressed in terms of integrals of the metric fields and can be applied in a straightforward way to known gravity dual theories. For example, anisotropic RG flows with AdS UV asymptotics were recently
constructed to study of the effect the confinement/deconfinemt phase
transitions and the inverse anisotropic catalysis effect \cite{Giataganas:2017koz}. A numerical analysis of \eq{derxexact},
could be applied to such vacuum gravity solutions to examine if there
is a need to constrain further the parametric space in those theories. Other theories with
anisotropic flows that our analysis can be applied include
\cite{Jain:2015txa,Arefeva:2018hyo,Mateos:2011ix,Giataganas:2013lga}.

It should be interesting to study the behavior of the
$c$-functions defined here at the quantum and topological anisotropic
 phase transitions. The entanglement entropy is an order parameter for
 such phase transitions and therefore the $c$-function itself is
 expected to show certain signals of discontinuity at the critical
 region.  Moreover, it would be very interesting to provide further
 evidence on our proposal by looking at the properties of the
 $c$-function in the weakly coupled anisotropic theories. The
 possibility that linear combinations of $c_x$ and $c_y$ may form a
 candidate for a holographic $c$-function cannot be excluded. In fact
 for theories with isotropic UV asymptotics the sum of the
 $c$-functions produce a invariant integral under interchanges
 $x\leftrightarrow y$, while for anisotropic UV dynamics the
 combination needs to be more involved.

We would also like to comment on further alternative applications of
our work.  In 3-dimensional isotropic CFTs, it has been shown that the free energy of the
theory on the $S^3$ coincides with the entanglement entropy of a
spherical surface, and therefore
it can be expressed via the entropic
formulation of the c-function \cite{Casini:2011kv,Ghosh:2018qtg}.
Generalization to more dimensions has also been found
\cite{Giombi:2014xxa,Kawano:2014moa}. It would be interesting to
extend the $F$-theorem for anisotropic flows using the findings of our
work.  Moreover, our study can be reformulated with the renormalized
entanglement entropy (REE) \cite{Liu:2012eea}, where all the potential
divergent terms along the RG flow have been removed and in certain
cases can play the role of the $F$-function. Furthermore, the
technical methods developed in our work may be applied to other non-local observables, like the heavy quark
observables.  As long as the observables are expressed holographically
in terms of the metric fields for general holographic backgrounds
(i.e. as in \cite{Giataganas:2012zy}), our methods can be applied to
study their
flow behavior along the RG trajectory.

\section*{Acknowledgements}

The authors acknowledge useful conversations with I. Papadimitriou.
C-S. Chu is supported by the grant 107-2119-M-007-014-MY3 from the Ministry of
Science and Technology of Taiwan.
D.G. research has been
funded by the Hellenic Foundation for Research and Innovation (HFRI)
and the General Secretariat for Research and Technology (GSRT), under
grant agreement No 2344.

%%%%%%%%%%%%%%%%%%%%%%%%%%%%%%%%%%%%%%%%%%%%%%%%%%%%%%%%%%%%%%%%%%%%%%%%%%%%%%%
%%%%%%%%%%%%%%%%%%%%%%%%%%%%%%%%%%%%%%%%%%%%%%%%%%%%%%%%%%%%%%%%%%%%%%%%%%%%%%%
%%%%%%%%%%%%%%%%%%%%%%%%%%%%%%%%%%%%%%%%%%%%%%%%%%%%%%%%%%%%%%%%%%%%%%%%%%%%%%%

\begin{appendices}

  \section{An Alternative Form for the Sufficient Conditions of Monotonicity}
  \la{app:a1}

A set of more restricting sufficient conditions of monotonicity are
derived by eliminating the second derivatives of \eq{derx} and
\eq{dery} using the null energy conditions to obtain
\bea\la{m11}
&&\frac{\partial c_x}{\pp r_m}= \frac{\b_x e^{k_m} l_x^{d_{x}-1} d_{x} }
       {4 G_N^{(d+2)}} k_m'\int_0^l dx \frac{1}{k'(r)^2}\prt{P_{+x} -
         \prt{B'(r) +A_1'(r)} k'(r)}~,\\\la{m22}
       &&\frac{\partial c_y}{\pp r_m}= \frac{\b_y e^{k_m} l_y^{d_{y}-1} d_{y}  }
       {4 G_N^{(d+2)}} k_m'\int_0^l dy \frac{1}{k'(r)^2}\prt{P_{+y} -
         \prt{B'(r) +A_2'(r)} k'(r)}~,\\\la{pplus}
&&P_{+x} := \frac{1}{d_x}k'(r)^2+d_1 A_1'(r)^2+d_2 A_2'(r)^2 + N_3 \ge 0~.
\eea
The conditions below are sufficient to guarantee monotonicity
\bea\la{strictcon}
k'(r)\ge 0 ~  \quad \mbox{and} \quad B'(r)+A_1'(r)\le 0~ \quad \mbox{and}
\quad B'(r)+A_2'(r)\le 0~.
\eea
We remark that the conditions \eq{strictcon} are sufficient to ensure a well behaved
monotonic flow and take simpler form compared to
\eq{sufcon1}-\eq{sufcon3}. However, by gaining on simplicity for
the sufficient conditions, they become more restrictive.

Let us note that in the case of AdS UV asymptotics, the expression of $P_{+x}$ simplifies to
\be
P_{+x}=-f_b'(r)  e^{k(r)/\prt{d_1+d_2}+B(r)}~,
\ee
where
\be
f_b(r):=k'(r) e^{-k(r)/\prt{d_1+d_2}-B(r)}~.
\ee
The NEC \eq{nec3} can be expressed in terms of  $f_b$ as
\be\la{eqfb}
f_b'(r)e^{k(r)/\prt{d_1+d_2}+B(r)}+\frac{2d_1^2 A_1'(r)^2+2 d_2^2 A_2'(r)+d_1 d_2
  \prt{A_1'(r) +A_2'(r)}^2}{d_1+d_2} \le 0~,
\ee
which implies that $f_b$ is a monotonically decreasing function. Therefore, an analysis along the lines of the section \ref{sec:uv} can be repeated here, which would lead to stricter UV boundary conditions  that ensure the $c$-function monotonicity.

\end{appendices}

\bibliographystyle{JHEP}

\end{document}